\documentclass[prl,superscriptaddress,twocolumn,showpacs,amsmath,amssymb]{revtex4}

\usepackage{epsfig}
\usepackage{dcolumn}
\usepackage{bm}
\usepackage{ulem}

\usepackage{graphicx}

\include{include}

\newcommand{\MetRaw}{\mbox{$E\kern-0.60em\raise0.10ex\hbox{/}_{T}^{\rm raw}$}}
\newcommand{\MetVec}{\mbox{$\vec E\kern-0.60em\raise0.10ex\hbox{/}_{T}$}}
\newcommand{\MetVecRaw}{\mbox{$\vec E\kern-0.60em\raise0.10ex\hbox{/}^{\rm raw}_{T}$}}
\newcommand{\Met}{\mbox{$E\kern-0.60em\raise0.10ex\hbox{/}_{T}$}}
\newcommand{\MetSpec}{\mbox{$E\kern-0.60em\raise0.10ex\hbox{/}_{T}^{spec}$}}
\newcommand{\MetRel}{\mbox{$E\kern-0.60em\raise0.10ex\hbox{/}_{T, rel}$}}

\newcommand{\WJ}{\mbox{$W+\textrm{1-jet}$}}

\begin{document}

\title{Measurement of the $W^+W^-$ Production Cross Section and Search for Anomalous $WW\gamma$ and $WWZ$ Couplings 
       in $p \bar p$ Collisions at $\sqrt{s} = 1.96$ TeV}
\date{\today}

\affiliation{Institute of Physics, Academia Sinica, Taipei, Taiwan 11529, Republic of China} 
\affiliation{Argonne National Laboratory, Argonne, Illinois 60439} 
\affiliation{University of Athens, 157 71 Athens, Greece} 
\affiliation{Institut de Fisica d'Altes Energies, Universitat Autonoma de Barcelona, E-08193, Bellaterra (Barcelona), Spain} 
\affiliation{Baylor University, Waco, Texas  76798} 
\affiliation{Istituto Nazionale di Fisica Nucleare Bologna, $^{cc}$University of Bologna, I-40127 Bologna, Italy} 
\affiliation{Brandeis University, Waltham, Massachusetts 02254} 
\affiliation{University of California, Davis, Davis, California  95616} 
\affiliation{University of California, Los Angeles, Los Angeles, California  90024} 
\affiliation{University of California, San Diego, La Jolla, California  92093} 
\affiliation{University of California, Santa Barbara, Santa Barbara, California 93106} 
\affiliation{Instituto de Fisica de Cantabria, CSIC-University of Cantabria, 39005 Santander, Spain} 
\affiliation{Carnegie Mellon University, Pittsburgh, PA  15213} 
\affiliation{Enrico Fermi Institute, University of Chicago, Chicago, Illinois 60637}
\affiliation{Comenius University, 842 48 Bratislava, Slovakia; Institute of Experimental Physics, 040 01 Kosice, Slovakia} 
\affiliation{Joint Institute for Nuclear Research, RU-141980 Dubna, Russia} 
\affiliation{Duke University, Durham, North Carolina  27708} 
\affiliation{Fermi National Accelerator Laboratory, Batavia, Illinois 60510} 
\affiliation{University of Florida, Gainesville, Florida  32611} 
\affiliation{Laboratori Nazionali di Frascati, Istituto Nazionale di Fisica Nucleare, I-00044 Frascati, Italy} 
\affiliation{University of Geneva, CH-1211 Geneva 4, Switzerland} 
\affiliation{Glasgow University, Glasgow G12 8QQ, United Kingdom} 
\affiliation{Harvard University, Cambridge, Massachusetts 02138} 
\affiliation{Division of High Energy Physics, Department of Physics, University of Helsinki and Helsinki Institute of Physics, FIN-00014, Helsinki, Finland} 
\affiliation{University of Illinois, Urbana, Illinois 61801} 
\affiliation{The Johns Hopkins University, Baltimore, Maryland 21218} 
\affiliation{Institut f\"{u}r Experimentelle Kernphysik, Karlsruhe Institute of Technology, D-76131 Karlsruhe, Germany} 
\affiliation{Center for High Energy Physics: Kyungpook National University, Daegu 702-701, Korea; Seoul National University, Seoul 151-742, Korea; Sungkyunkwan University, Suwon 440-746, Korea; Korea Institute of Science and Technology Information, Daejeon 305-806, Korea; Chonnam National University, Gwangju 500-757, Korea; Chonbuk National University, Jeonju 561-756, Korea} 
\affiliation{Ernest Orlando Lawrence Berkeley National Laboratory, Berkeley, California 94720} 
\affiliation{University of Liverpool, Liverpool L69 7ZE, United Kingdom} 
\affiliation{University College London, London WC1E 6BT, United Kingdom} 
\affiliation{Centro de Investigaciones Energeticas Medioambientales y Tecnologicas, E-28040 Madrid, Spain} 
\affiliation{Massachusetts Institute of Technology, Cambridge, Massachusetts  02139} 
\affiliation{Institute of Particle Physics: McGill University, Montr\'{e}al, Qu\'{e}bec, Canada H3A~2T8; Simon Fraser University, Burnaby, British Columbia, Canada V5A~1S6; University of Toronto, Toronto, Ontario, Canada M5S~1A7; and TRIUMF, Vancouver, British Columbia, Canada V6T~2A3} 
\affiliation{University of Michigan, Ann Arbor, Michigan 48109} 
\affiliation{Michigan State University, East Lansing, Michigan  48824}
\affiliation{Institution for Theoretical and Experimental Physics, ITEP, Moscow 117259, Russia} 
\affiliation{University of New Mexico, Albuquerque, New Mexico 87131} 
\affiliation{Northwestern University, Evanston, Illinois  60208} 
\affiliation{The Ohio State University, Columbus, Ohio  43210} 
\affiliation{Okayama University, Okayama 700-8530, Japan} 
\affiliation{Osaka City University, Osaka 588, Japan} 
\affiliation{University of Oxford, Oxford OX1 3RH, United Kingdom} 
\affiliation{Istituto Nazionale di Fisica Nucleare, Sezione di Padova-Trento, $^{dd}$University of Padova, I-35131 Padova, Italy} 
\affiliation{LPNHE, Universite Pierre et Marie Curie/IN2P3-CNRS, UMR7585, Paris, F-75252 France} 
\affiliation{University of Pennsylvania, Philadelphia, Pennsylvania 19104}
\affiliation{Istituto Nazionale di Fisica Nucleare Pisa, $^{ee}$University of Pisa, $^{ff}$University of Siena and $^{gg}$Scuola Normale Superiore, I-56127 Pisa, Italy} 
\affiliation{University of Pittsburgh, Pittsburgh, Pennsylvania 15260} 
\affiliation{Purdue University, West Lafayette, Indiana 47907} 
\affiliation{University of Rochester, Rochester, New York 14627} 
\affiliation{The Rockefeller University, New York, New York 10021} 
\affiliation{Istituto Nazionale di Fisica Nucleare, Sezione di Roma 1, $^{hh}$Sapienza Universit\`{a} di Roma, I-00185 Roma, Italy} 

\affiliation{Rutgers University, Piscataway, New Jersey 08855} 
\affiliation{Texas A\&M University, College Station, Texas 77843} 
\affiliation{Istituto Nazionale di Fisica Nucleare Trieste/Udine, I-34100 Trieste, $^{ii}$University of Trieste/Udine, I-33100 Udine, Italy} 
\affiliation{University of Tsukuba, Tsukuba, Ibaraki 305, Japan} 
\affiliation{Tufts University, Medford, Massachusetts 02155} 
\affiliation{Waseda University, Tokyo 169, Japan} 
\affiliation{Wayne State University, Detroit, Michigan  48201} 
\affiliation{University of Wisconsin, Madison, Wisconsin 53706} 
\affiliation{Yale University, New Haven, Connecticut 06520} 
\author{T.~Aaltonen}
\affiliation{Division of High Energy Physics, Department of Physics, University of Helsinki and Helsinki Institute of Physics, FIN-00014, Helsinki, Finland}
\author{J.~Adelman}
\affiliation{Enrico Fermi Institute, University of Chicago, Chicago, Illinois 60637}
\author{B.~\'{A}lvarez~Gonz\'{a}lez$^v$}
\affiliation{Instituto de Fisica de Cantabria, CSIC-University of Cantabria, 39005 Santander, Spain}
\author{S.~Amerio$^{dd}$}
\affiliation{Istituto Nazionale di Fisica Nucleare, Sezione di Padova-Trento, $^{dd}$University of Padova, I-35131 Padova, Italy} 

\author{D.~Amidei}
\affiliation{University of Michigan, Ann Arbor, Michigan 48109}
\author{A.~Anastassov}
\affiliation{Northwestern University, Evanston, Illinois  60208}
\author{A.~Annovi}
\affiliation{Laboratori Nazionali di Frascati, Istituto Nazionale di Fisica Nucleare, I-00044 Frascati, Italy}
\author{J.~Antos}
\affiliation{Comenius University, 842 48 Bratislava, Slovakia; Institute of Experimental Physics, 040 01 Kosice, Slovakia}
\author{G.~Apollinari}
\affiliation{Fermi National Accelerator Laboratory, Batavia, Illinois 60510}
\author{A.~Apresyan}
\affiliation{Purdue University, West Lafayette, Indiana 47907}
% ...

\collaboration{CDF Collaboration\footnote{With visitors from $^a$University of Massachusetts Amherst, Amherst, Massachusetts 01003,
$^b$Universiteit Antwerpen, B-2610 Antwerp, Belgium, 
$^c$University of Bristol, Bristol BS8 1TL, United Kingdom,
$^d$Chinese Academy of Sciences, Beijing 100864, China, 
$^e$Istituto Nazionale di Fisica Nucleare, Sezione di Cagliari, 09042 Monserrato (Cagliari), Italy,
$^f$University of California Irvine, Irvine, CA  92697, 
$^g$University of California Santa Cruz, Santa Cruz, CA  95064, 
$^h$Cornell University, Ithaca, NY  14853, 
$^i$University of Cyprus, Nicosia CY-1678, Cyprus, 
$^j$University College Dublin, Dublin 4, Ireland,
$^k$University of Edinburgh, Edinburgh EH9 3JZ, United Kingdom, 
$^l$University of Fukui, Fukui City, Fukui Prefecture, Japan 910-0017
$^m$Kinki University, Higashi-Osaka City, Japan 577-8502
$^n$Universidad Iberoamericana, Mexico D.F., Mexico,
$^o$University of Iowa, Iowa City, IA  52242,
$^p$Kansas State University, Manhattan, KS 66506
$^q$Queen Mary, University of London, London, E1 4NS, England,
$^r$University of Manchester, Manchester M13 9PL, England,
$^s$Muons, Inc., Batavia, IL 60510, 
$^t$Nagasaki Institute of Applied Science, Nagasaki, Japan, 
$^u$University of Notre Dame, Notre Dame, IN 46556,
$^v$University de Oviedo, E-33007 Oviedo, Spain, 
$^w$Texas Tech University, Lubbock, TX  79609, 
$^x$IFIC(CSIC-Universitat de Valencia), 56071 Valencia, Spain,
$^y$Universidad Tecnica Federico Santa Maria, 110v Valparaiso, Chile,
$^z$University of Virginia, Charlottesville, VA  22906
$^{aa}$Bergische Universit\"at Wuppertal, 42097 Wuppertal, Germany,
$^{bb}$Yarmouk University, Irbid 211-63, Jordan
$^{jj}$On leave from J.~Stefan Institute, Ljubljana, Slovenia, 
}}
\noaffiliation


\begin{abstract} 
This Letter describes the current most precise measurement of the $W$ boson pair production cross section  and most sensitive test of anomalous $WW\gamma$ and $WWZ$ couplings in $p \bar p$ collisions at a center-of-mass energy of 1.96 TeV. The $WW$ candidates are reconstructed from decays containing two charged leptons and two neutrinos,  where the charged leptons are either electrons or muons. Using data collected by the CDF II detector from 3.6 fb$^{-1}$ of integrated luminosity, a total of 654 candidate events are observed with an expected background  contribution of $320 \pm 47$ events. The measured total cross section is $\sigma (p \bar p \to W^+ W^- + X) = 12.1 \pm 0.9 \textrm{ (stat)} \;^{+1.6}_{-1.4} \; \textrm{(syst)}$ pb, which is in good agreement with the standard model prediction. The same data sample is used to place constraints on anomalous $WW\gamma$ and $WWZ$ couplings. 

\end{abstract}


\maketitle

\normalsize

The measurement of $W$ boson pair production is an important test of the standard model 
(SM) of particle physics.  This process is also an essential background
to understand for Higgs boson searches 
at particle colliders.  Next-to-leading order (NLO) calculations of $W^+W^-$ production in $p \bar p$ 
collisions at $\sqrt{s} = 1.96$ TeV predict a cross section of 
$\sigma^{NLO}(p \bar p \rightarrow W^+W^-) = 11.7 \pm 0.7$ pb~\cite{MCFM,MCatNLO}.
The presence of anomalous $WW\gamma$ and $WWZ$ triple gauge boson couplings (TGC)~\cite{ATGC}
could be indications of new physics at a higher mass scale, and would lead to rates for $W^+W^-$ 
production or kinematic distributions that differ from those predicted by the SM.  

This Letter reports a measurement of the $W^+W^-$ production cross section 
and limits on anomalous TGCs using a final state consisting of two oppositely charged 
leptons and two neutrinos in $p \bar p$ collision data collected by the CDF II detector
from 3.6 fb$^{-1}$ of integrated luminosity.
First evidence for $W$ boson pair production was reported by CDF using Tevatron Run I 
data~\cite{CDFEVIDENCE}.  This process was later measured 
with greater significance by CDF and D0 using 184 pb$^{-1}$ and 224-252 pb$^{-1}$ 
respectively of integrated luminosity from Run II of the Tevatron~\cite{CDFWW,D0WW}. 
Recently, the D0 collaboration measured the $W^+W^-$ cross section with a precision of 20\% 
using 1.0 fb$^{-1}$ of integrated luminosity from Run II~\cite{newD0WW}.
Limits on anomalous TGCs have previously been reported by LEP
experiments and the CDF and D0 collaborations \cite{newD0WW,existingTGCLimits}.

The cross section measurement 
uses a matrix element method in which the probability for each event to have been
produced by several relevant SM processes is calculated.
A likelihood ratio ($LR$) is formed from these probabilities.
The predicted shapes and normalizations of the signal and background $LR$ distributions are 
used to extract the SM $W^+W^-$ production cross section via a maximum-likelihood fit to the 
$LR$ distribution observed in data. 
In general, the presence of anomalous TGCs will increase the number of events containing
leptons with very high values of momentum transverse to the beam direction, as compared to
SM expectations.
Limits on anomalous TGCs are determined from the shape and normalization of the transverse 
momentum spectrum constructed from the highest transverse momentum lepton in the event,
referred to as the leading lepton.
The results are reported in the HISZ scheme, where three parameters, $\lambda_Z$, $g_1^Z$, 
and $\kappa_\gamma$, are used to describe all dimension-six operators which are Lorentz and
$SU(2)_L \otimes U(1)_Y$ invariant and conserve $C$ and $P$ separately \cite{HISZ}.
In the SM, $\lambda_Z = 0$ and $g_1^Z = \kappa_\gamma = 1$.
In this Letter, $\Delta g_1^Z$ and $\Delta \kappa_\gamma$ are used to denote the 
deviation of the $g_1^Z$ and $\kappa_\gamma$ parameters from their SM values.
The non-SM values of the parameters $\lambda_Z$, $g_1^Z$, and $\kappa_\gamma$ are 
functions of the invariant mass of the $W^+W^-$ system, $\sqrt{\hat{s}}$.  These results 
probe a larger range of values of $\sqrt{\hat{s}}$ and thus
may only be qualitatively compared to the results from LEP, which were below $\sqrt{\hat{s}} = 209$ GeV.
For hadron collisions, a dipole form factor for an arbitrary coupling 
$\alpha(\hat{s})=\frac{\alpha_0}{(1+\hat{s}/\Lambda)^2}$ \cite{HISZ} is 
introduced to turn off the coupling at large $\sqrt{\hat{s}}$ and avoid a violation of unitarity.
The form factor scale $\Lambda$ is the scale of new physics.

In the CDF II detector~\cite{CDFII}, a particle's direction
is characterized by the azimuthal angle $\phi$ and the pseudorapidity
$\eta = -\ln [ \tan (\theta/2) ]$, where $\theta$ is the 
polar angle measured from the proton beam direction.  The transverse energy $E_T$ 
is defined as $E \sin \theta$, where $E$ is the energy in the 
calorimeter towers associated with a cluster of energy deposition.  
Transverse momentum, $p_T$, is the track momentum component transverse to the beam line.
The magnitude of the $p_T$ for an electron is scaled according to the energy measured in the
calorimeter.
The missing transverse energy vector, $\MetVec$, is defined as $-\sum_{i} E_T^i ~ \hat{n}_T^i$, where 
$\hat{n}_T^i$ is the unit vector in the transverse plane pointing from the interaction point to the 
energy deposition in calorimeter tower $i$.  This is corrected for the $p_T$ of muons, which do 
not deposit all of their energy in the calorimeter, and tracks which point to uninstrumented 
regions in the calorimeter.  The scalar $\Met$ is defined as $|\MetVec|$.
Strongly interacting partons produced in the $p \bar p$ collision undergo fragmentation
that results in highly collimated jets of hadronic particles.  Jet candidates are
reconstructed using the calorimeter 
and are required to have $E_T > 15$ GeV and $|\eta| < 2.5$.
Isolated lepton candidates are accepted out to an $|\eta|$ of 2.0 for electron candidates and 
$|\eta|$ of 1.0 for muon candidates.

The experimental signature for the decay $W^+W^- \to \ell^+\nu \ell^-\bar{\nu}$ is two reconstructed leptons with opposite
charge and $\Met$ from the neutrinos which escape undetected. 
In this Letter, $\ell$ refers to an electron or muon.  
Additional signal acceptance ($\sim$12\%) is obtained from cases where one or both $W$ 
bosons decay to a tau lepton which subsequently decays to an electron or muon.
There are several SM processes which result in a similar 
final state to $W^+ W^-$ and are therefore backgrounds in this measurement.  These are other diboson 
production ($WZ$, $ZZ$) and top-quark pair production ($t \bar t$).
It is also possible to observe apparent $\Met$ arising from the mismeasurement 
of lepton energy, lepton momentum, or the hadronic part of the final state.  
Drell-Yan ($Z/\gamma^* \to \ell^+ \ell^-$) events have no neutrinos in the final 
state, but due to large production rates enter the $W^+ W^-$ candidate sample via mismeasurements.
A third source of background is events in which a final-state particle is misidentified.
These are $W+$jets and $W\gamma$ production, where the $W$ boson decays leptonically and a jet is reconstructed 
as a lepton candidate or the $\gamma$ converts in the detector material and is reconstructed
as an electron.

Events containing two oppositely charged lepton candidates are selected from the data sample.
The online event triggering and selection of lepton candidates are identical to those used in the
search for SM Higgs bosons decaying to two $W$ bosons at CDF~\cite{CDFHWW}.
%
The leading-lepton $p_T$ is required to be above 20 GeV/$c$ to satisfy the trigger requirements, 
while the second lepton is allowed to have a $p_T$ as low as $10$ GeV/$c$.  
The requirement is also made that events contain no jet candidates, 
which significantly reduces $t \bar t$ background.
A variant of $\Met$ used in selecting candidate events is defined as
$\MetRel = \Met \sin \Delta \phi(\Met, \ell)$ when $\Delta \phi(\Met, \ell) \leq \frac{\pi}{2}$,
where $\Delta \phi(\Met, \ell)$ is the azimuthal separation between the $\MetVec$ and the 
momentum vector of the nearest lepton candidate.  
If $\Delta \phi(\Met, \ell) > \frac{\pi}{2}$, then $\MetRel = \Met$.
The $\MetRel$ variable is designed to reject events where the apparent $\Met$
arises from the mismeasurement of lepton energy or momentum, and
is required to be above 25 GeV to reduce the otherwise large Drell-Yan contamination.  
This requirement is lowered to 15 GeV for electron-muon events where contributions from Drell-Yan 
are inherently smaller.
The $W\gamma$ and heavy-flavor ($J/\psi, \Upsilon$) backgrounds are reduced by requiring
that the invariant mass of the lepton pair be greater than 16 GeV/$c^2$.

With the exception of the \WJ~background, the acceptance
and kinematic properties of the signal and background
processes are determined by simulation.  Events from $W^+W^-$ are simulated at
NLO using the {\sc mc@nlo} generator~\cite{MCatNLO}.  The $t \bar t$, 
$WZ$, $ZZ$, and Drell-Yan backgrounds are simulated with the {\sc pythia} generator~\cite{PYTHIA}.  
The $W\gamma$ background is determined using the generator 
described in Ref.~\cite{BAUR}.
The response of the CDF II detector is modelled with a
{\sc geant-3}-based simulation~\cite{GEANT}.  
The expected yields for each process are normalized to the cross sections calculated at
partial next-to next-to-leading order ($t \bar t$~\cite{ttbarXsec}),
NLO ($W^+W^-$~\cite{MCFM, MCatNLO}, $WZ$ and $ZZ$~\cite{MCFM}), or
leading-order with estimated higher-order corrections ($W\gamma$~\cite{BAUR} and Drell-Yan~\cite{DY}).
Efficiency corrections for the simulated detector response to lepton candidates
are determined using samples of observed $Z \to \ell^+ \ell^-$ events.
%
The \WJ~background is calculated using the
probability, measured in independent jet-triggered data samples, that a hadronic jet will be reconstructed as a
lepton candidate.
These probabilities are applied to the jet in the \WJ~data sample to estimate the 
number of such events which will pass the full lepton identification and signal selection 
criteria.  The expected signal and background contributions are given in Table~\ref{tbl:WWExpected}
along with the observed number of events.

\begin{table}
\centering
\caption{Expected number of signal ($W^+W^-$) and background events along with the
	total number of expected and observed events in the data.  Uncertainties 
	include all systematic uncertainties described in the text.}
\label{tbl:WWExpected}
\begin{tabular}{lrcr}\hline\hline
Process           &   \multicolumn{3}{c}{Events}   \\\hline
$Z/\gamma^*$ (Drell-Yan)     &          79.8 &$\pm$&    18.4  \\
$WZ$              &          13.8 &$\pm$&     1.9  \\
$W\gamma$         &          91.7 &$\pm$&    24.8  \\
\WJ	              &         112.7 &$\pm$&    31.2  \\
$ZZ$              &          20.7 &$\pm$&     2.8  \\
$t \bar t$        &           1.3 &$\pm$&     0.2  \\\hline
Total Background  &         320.0 &$\pm$&    46.8  \\\hline
$W^+W^-$          &         317.6 &$\pm$&    43.8  \\\hline
Total Expected    &         637.6 &$\pm$&    73.0  \\\hline
Data              &  \multicolumn{3}{c}{654}       \\\hline\hline
\end{tabular}
\end{table}

The dominant systematic uncertainties on the estimated contributions come from the luminosity 
measurement (6\%)~\cite{LUMI} and the simulated acceptances of the signal and background processes.
The acceptance uncertainty due to the parton distribution function modeling
ranges from 1.9\% to 4.1\% for the different processes.
A 10\% uncertainty is assigned to all simulated processes
for the kinematic differences between leading-order and higher-order calculations,
based on the difference in acceptance of $W^+W^-$ events simulated at leading-order
and NLO using the {\sc pythia} and {\sc mc@nlo} generators respectively.
This uncertainty is reduced to 5\% for the $W^+W^-$ signal, which is simulated at NLO.
The cross section uncertainties are 6\% on diboson production, 10\% on $t \bar t$
and $W\gamma$ production, and 5\% on Drell-Yan production.  
A 21\% uncertainty is included for the Drell-Yan background to account for the mismodeling of 
$\Met$ and jet production rates.
Systematic uncertainties of 20\% and 27\% are
assigned to the $W\gamma$ and \WJ~background estimates, respectively,
due to uncertainties in the modeling
of the photon conversions and misidentification of a jet as a lepton.
%
Uncertainties on the modeling of jets accounts for 2\% to 4\% and 
lepton identification and trigger efficiencies range from 1\% to 7\%.

For each event passing the signal selection criteria, four matrix-element-based event
probabilities are calculated corresponding to the production and decay 
processes $W^+ W^- \rightarrow \ell^+ \nu \ell^- \bar \nu$, $ZZ \rightarrow \ell^+ \ell^- \nu \bar \nu$,
$\WJ \rightarrow \ell \nu +$ 1-jet, and $W\gamma \rightarrow \ell \nu + \gamma$.
In the latter two processes, the jet or $\gamma$ is assumed to have been
reconstructed as a charged lepton candidate.  The event probability 
for a process $X$ is given by
\begin{equation}
\label{eqn:MEProb}
P_{X}(\vec x) = \frac{1}{\langle \sigma \rangle} \int \frac{d \sigma(\vec y) }{d \vec y} \; \epsilon(\vec y) \; G(\vec x, \vec y) \; d\vec y
\end{equation}
where $\vec x$ represents the observed lepton momenta and $\MetVec$ vectors, 
$G(\vec x, \vec y)$ is a transfer function representing the detector resolution, and
$\epsilon(\vec y)$ is an efficiency function parametrized by $\eta$ which quantifies 
the probability for a particle to be reconstructed as a lepton.  
The differential cross section $\frac{d \sigma(\vec y) }{d \vec y}$ is calculated using 
leading-order matrix elements from the {\sc mcfm} program~\cite{MCFM} and integrated over all 
possible true values of the final state particle 4-vectors $\vec y$.  The normalization factor
$\langle \sigma \rangle$ is determined from the leading-order cross section and
detector acceptance for each process.
These event probabilities are combined into a likelihood ratio
\begin{equation}
\label{eqn:LR}
LR_{WW} = \frac{ P_{WW} }{ P_{WW} + \sum_{j} k_j P_j},
\end{equation}
where $j = \{ZZ, \WJ, W\gamma\}$ and  $k_j$ is the relative fraction of the expected number 
of events for the $j$-th process such that $\sum_j k_j = 1$.  The templates of the $LR_{WW}$
distribution are created for signal and each background process given in Table~\ref{tbl:WWExpected}.

\begin{figure}
\centering
\includegraphics[width=\linewidth]{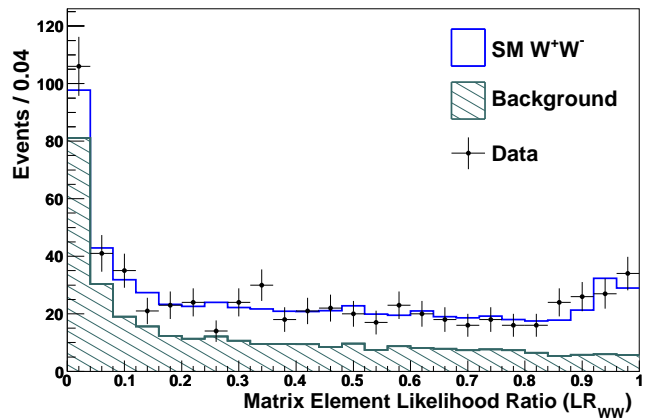}
\caption{The $LR_{WW}$ distributions for the signal ($W^+W^-$) and background processes 
  after a maximum likelihood fit to the data.}
\label{fig:WWFit}
\end{figure}

A binned maximum likelihood is used to extract the $W^+W^-$ production cross section from 
the shape and normalization of the $LR_{WW}$ templates.
The likelihood is formed from the Poisson probabilities of observing $n_i$ events in the $i$-th 
bin when $\mu_i$ are expected.  Variations corresponding to the systematic uncertainties
described previously are included as normalization parameters for signal and 
background, constrained by Gaussian terms.
The likelihood is given by
\begin{equation}
\label{eqn:Likelihood}
\mathcal{L} = \left(\prod_i \frac{\mu_i^{n_i} e^{-\mu_i}}{n_i!} \right) \cdot \prod_c e^{-\frac{S_c^2}{2}}
\end{equation}
where
\begin{equation}
\label{eqn:MuI}
\mu_i = \sum_k \alpha_k \left[ \prod_c (1 + f^c_k S_c ) \right]  (N^{Exp}_k)_i ,
\end{equation}
$f^c_k$ is the fractional uncertainty for the process $k$ due to the systematic $c$,
and $S_c$ is a floating parameter associated with the systematic uncertainty $c$.  
The correlations of systematic uncertainties between processes are accounted for in the definition of $\mu_i$.
The expected number of events from process $k$ in
the $i$-th bin is given by $(N^{Exp}_k)_i$.
The parameter $\alpha_k$ is an overall normalization parameter for process $k$ and 
is fixed to unity for all processes other than $W^+W^-$, for which it is freely floating.  
The likelihood is maximized with respect to the 
systematic parameters $S_c$ and $\alpha_{WW}$ using the {\sc minuit} program~\cite{MINUIT}. 
The $W^+W^-$ cross section is then given by the fitted value of $\alpha_{WW}$ multiplied by $\sigma^{NLO} (p \bar p \to W^+W^-)$.
%
This method gives a measured value for the $W^+W^-$ production cross section of
$\sigma (p \bar p \to W^+ W^- + X) = 12.1 \pm 0.9 \textrm{ (stat)} \;^{+1.6}_{-1.4} \; \textrm{(syst)}$ pb.
The fit to the data of the signal and sum of the individually fitted background templates is shown
in Fig.~\ref{fig:WWFit}.

\begin{figure}
\centering
\includegraphics[width=\linewidth]{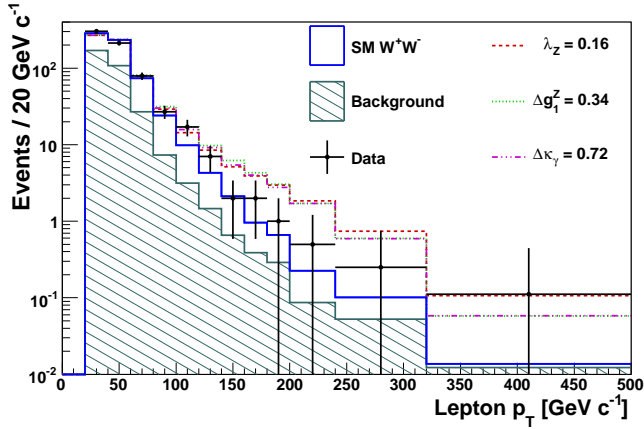}
\caption{Leading-lepton $p_T$ distribution for data compared to the SM expectation.
         Also shown is how the expectation
         would be modified by anomalous couplings near the observed limits.}
\label{fig:TGC}
\end{figure}

The likelihood of the observed leading-lepton $p_T$ 
distribution is used to set limits on anomalous TGC
values. 
The robustness of the leading-lepton $p_T$ distribution has been verified using the
same lepton selection in several non-overlapping final state kinematic regions.
%
The response of the detector to events with different coupling constants is simulated 
for six points in the parameter space near the existing limits~\cite{existingTGCLimits}.
The efficiency 
multiplied by acceptance as a function of the leading-lepton $p_T$ is taken to be the average
of the values measured in these samples.  The uncertainty is taken to be the maximum variation 
among these samples and 
ranges from 7\% at low $p_T$ to 50\% at high $p_T$.
This $p_T$-dependent efficiency is applied to the NLO generator-level
distributions produced by the {\sc mcfm} program~\cite{MCFM} to predict the leading-lepton $p_T$ spectrum 
for the coupling values considered, as shown in Fig.~\ref{fig:TGC}. 

Each of the likelihoods ${\cal L}(\lambda_Z)$, ${\cal L}(\Delta g_1^Z)$, and ${\cal L}(\Delta \kappa_\gamma)$
are computed as the product over all bins in the leading-lepton $p_T$ distribution
of the Poisson probability of each bin given the model,
and 95\% confidence levels are set where $(-2\ln{\cal L})-(-2\ln{\cal L}_{min})=(1.96)^2$.
The systematic uncertainties include all those described for the $W^+ W^-$ cross section and the
additional $p_T$-dependent uncertainty on the efficiency described previously.
Systematic uncertainties are implemented by simultaneously
applying all variations which reduce the sensitivity.
The observed 95\% confidence limits, shown in Table 
\ref{tab:TGC}, are weaker than expected. The probability of observing these limits
in the presence of only standard model $W^+ W^-$ production ranges from 7.1\% to 7.6\%
depending on the coupling constants ($\lambda_Z, g_1^Z,\kappa_\gamma$).

\begin{table}
\caption{Expected and observed limits on anomalous TGCs.  For each coupling limit set,
	the two other couplings are fixed at their SM values.  Values of the 
	couplings outside of the given observed range are excluded at the 95\% confidence level.}
\label{tab:TGC}
\begin{tabular}{llcccc}
\hline \hline
&&$ \Lambda$ (TeV)  &   $\lambda_Z$ &  $\Delta g_1^Z$  &   $\Delta \kappa_\gamma$ \\
\hline
Expected & ~~ & 1.5     & (-0.05,0.07) & (-0.09,0.17) & (-0.23,0.31) \\
Observed & ~~ & 1.5     & (-0.16,0.16) & (-0.24,0.34) & (-0.63,0.72) \\
\hline 	 
Expected & ~~ & 2.0     & (-0.05,0.06) & (-0.08,0.15) & (-0.20,0.27) \\
Observed & ~~ & 2.0     & (-0.14,0.15) & (-0.22,0.30) & (-0.57,0.65) \\
\hline \hline
\end{tabular}
\end{table}

In summary, the $W^+W^-$ production cross section has been measured in $p \bar p$ collisions 
at $\sqrt{s} = 1.96$ TeV from reconstructed events in the dilepton final state using a likelihood ratio formed from 
matrix-element-based event probabilities.
This result is the most precise measurement at this energy with an 
overall uncertainty of less than 15\%.
The same event sample is also used to
 perform the most sensitive probe to date at this energy 
of anomalous $WWZ$ and $WW\gamma$ couplings.
The leading-lepton $p_T$ distribution
of the sample is found to be in moderate agreement with the SM expectation and used to place
limits on anomalous triple gauge couplings.

We thank the Fermilab staff and the technical staffs of the participating institutions for their 
vital contributions. This work was supported by the U.S. Department of Energy and National Science 
Foundation; the Italian Istituto Nazionale di Fisica Nucleare; the Ministry of Education, Culture, 
Sports, Science and Technology of Japan; the Natural Sciences and Engineering Research Council of 
Canada; the National Science Council of the Republic of China; the Swiss National Science Foundation; 
the A.P. Sloan Foundation; the Bundesministerium f\"ur Bildung und Forschung, Germany; the World 
Class University Program, the National Research Foundation of Korea; the Science and Technology 
Facilities Council and the Royal Society, UK; the Institut National de Physique Nucleaire et Physique 
des Particules/CNRS; the Russian Foundation for Basic Research; the Ministerio de Ciencia e 
Innovaci\'{o}n, and Programa Consolider-Ingenio 2010, Spain; the Slovak R\&D Agency; and the Academy 
of Finland.


\end{document}